\RequirePackage{amsmath}
\documentclass[12pt]{iopart}
\newcommand{\BPEX}{{\it Biomed. Phys. Eng. Express}\xspace}

\usepackage{etoolbox}
\providetoggle{estiloharvard}
\providetoggle{forpublication}
\providetoggle{forarxiv}

\settoggle{estiloharvard}{false}
\settoggle{forpublication}{false}
\settoggle{forarxiv}{true}

\iftoggle{estiloharvard}{
\usepackage[abbr,dcu]{harvard}

\usepackage{hyperref}
\usepackage{verbatim}
}{
\usepackage{hyperref}
\usepackage[numbers,sort&compress]{natbib}
\def\newblock{\ }
\bibliographystyle{vancouver}
}

\usepackage{iopams}  

\usepackage[all]{hypcap}    

\usepackage[edtable,mathlines]{lineno}

\usepackage{xspace}
\usepackage[nolist]{acronym}

\newcommand*{\eg}{e.g.\@\xspace}
\newcommand*{\ie}{i.e.\@\xspace}
\newcommand*{\cf}{cf.\@\xspace}

\newcommand\refig[1]{fig.~\ref{#1}}

\newcommand{\tlow}{\textsubscript}

\newcommand{\code}[1]{\texttt{#1}}

\newcommand{\mystrut}{\rule[-0.8mm]{0pt}{3.5mm}}
\newcommand{\lstrut}{\vphantom{y}}

\usepackage{enumitem}

\usepackage{tikz}

\usetikzlibrary{shapes,arrows.meta,calc,fit,backgrounds,shapes.multipart,positioning,shadows}
\tikzset{box/.style={draw, rectangle, rounded corners, thick, node 
distance=7em, 
text width=6em, text centered, minimum height=3.5em}}

\tikzset{block/.style={rectangle split, draw, rectangle split parts=2,
text width=12em, text centered, rounded corners, minimum height=4em},
grnblock/.style={rectangle, draw, fill=green!20, text width=10em, text centered, rounded corners, minimum height=4em}, 
whtblock/.style={rectangle, draw, fill=white!20, text width=10em, text centered, minimum height=2em},    
line/.style={draw, -{Latex[length=2mm,width=1mm]}},
cloud/.style={draw, ellipse,fill=white!20, node distance=3cm,    minimum height=4em},  
container/.style={draw, rectangle,dashed,inner sep=0.28cm, rounded
corners,fill=yellow!20,minimum height=4cm}}

\makeatletter 
\tikzset{
  use path/.code={\pgfsyssoftpath@setcurrentpath{#1}}
}
\makeatother 
\usepackage{media9}

\makeatletter
\newcommand\notsotiny{\@setfontsize\notsotiny\@vipt\@viipt}
\makeatother
\usepackage{eso-pic}
\iftoggle{forarxiv}{
\AddToShipoutPictureBG{
  \AtPageLowerLeft{
    \raisebox{1.25\baselineskip}{\makebox[\paperwidth]{\begin{minipage}{8.5in}\centering
\notsotiny
      This Accepted Manuscript is available for reuse under a CC BY-NC-ND licence after the 12 month embargo period provided that all the terms of the licence are adhered to.
    \end{minipage}}}
  }
  \AtPageUpperLeft{
    \raisebox{-1.25\baselineskip}{\makebox[\paperwidth]{\begin{minipage}{8.5in}\centering
\notsotiny
      Citation information: F Hueso-Gonz\'alez et al 2020 \BPEX \textbf{6} 055013 \href{https://doi.org/10.1088/2057-1976/aba442}{https://doi.org/10.1088/2057-1976/aba442}
    \end{minipage}}}
  }
}
}{} 
\begin{acronym}[KVI-CART]
\acro{1D}{one-dimensional}
\acro{2D}{two-dimensional}
\acro{3D}{three-dimensional}
\acro{ADC}{Analog to Digital Converter}
\acro{AGOR}{Accélérateur Groningen ORsay}
\acro{API}{application programming interface}
\acro{BbCc}{BGO block Compton camera}
\acro{BGO}{Bismuth Germanium Oxide - Bi\tlow{4}Ge\tlow{3}O\tlow{12}}
\acro{BP}{Back Projection}
\acro{C230}{Cyclone\textsuperscript{\textregistered}\,230}
\acro{Cc}{Compton camera}
\acro{CERN}{Conseil Européen pour la Recherche Nucléaire}
\acro{CFD}{Constant Fraction Discriminator}
\acro{CPGM}{Coaxial Prompt Gamma-ray Monitoring}
\acro{CT}{computed tomography}
\acro{CTR}{Coincidence Time Resolution}
\acro{CSDA}{Continuous Slowing Down Approximation}
\acro{CZT}{Cadmium zinc telluride - CdZnTe}
\acro{DCEU}{Digital Counter Electronic Unit}
\acro{DDR3}{Double Data Rate type 3}
\acro{DSC}{dice similarity coefficient}
\acro{ELBE}{Electron Linear accelerator for beams with high Brilliance and low Emittance}
\acro{EPT}{Electron-Positron Tracking}
\acro{FP}{False Positive}
\acro{FPGA}{Field Programmable Gate Array}
\acro{FWHM}{Full Width at Half Maximum}
\acro{GEVI}{Gamma Electron Vertex Imaging}
\acro{GPIB}{General Purpose Interface Bus}
\acro{GUI}{graphical user interface}
\acro{GSO}{Cerium-doped gadolinium oxyorthosilicate - Gd\tlow{2}SiO\tlow{5}:Ce}
\acro{HV}{High Voltage}
\acro{KVI-CART}{Kernfysisch Versneller Instituut - Center for Advanced Radiation Technology}
\acro{LINAC}{Linear Accelerator}
\acro{LLUMC}{Loma Linda University Medical Center}
\acro{LSO}{Cerium-doped lutetium oxyorthosilicate - Lu\tlow{2}SiO\tlow{5}:Ce}
\acro{MAW}{Moving Average Window}
\acro{MC}{Monte Carlo}
\acro{MGH}{Massachusetts General Hospital}
\acro{MLEM}{Maximum-Likelihood Expectation-Maximisation}
\acro{MOSFET}{Metal-Oxide Semiconductor Field-Effect Transistor}
\acro{MRI}{magnetic resonance imaging}
\acro{NDA}{Non-Disclosure Agreement}
\acro{NIM}{Nuclear Instrumentation Module}
\acro{NPDF}{Normalised Probability Density Function}
\acro{OAR}{Organs At Risk}
\acro{PA}{Preamplifier}
\acro{PBS}{Pencil Beam Scanning}
\acro{PCIe}{Peripheral Component Interconnect Express}
\acro{PET}{Positron Emission Tomography}
\acro{PGI}{Prompt Gamma-ray Imaging}
\acro{PGPI}{Prompt Gamma-ray Peak Integration}
\acro{PGS}{Prompt Gamma-ray Spectroscopy}
\acro{PGT}{Prompt Gamma-ray Timing}
\acro{PLY}{Polygon File Format}
\acro{PMMA}{Polymethyl Methacrylate - [C\tlow{5}O\tlow{2}H\tlow{8}]\tlow{n}}
\acro{PMT}{Photomultiplier Tube}
\acro{QDC}{Charge to Digital Converter}
\acro{RBNS}{Recursive Bisection Neutron Subtraction}
\acro{RF}{Radio Frequency}
\acro{ROC}{Receiver Operating Characteristic}
\acro{ROI}{region of interest}
\acro{SDEP}{Single and Double Escape Peaks}
\acro{simBox}{simple Box}
\acro{SNIP}{Sensitive Nonlinear Iterative clipping Peak}
\acro{SOBP}{Spread Out Bragg Peak}
\acro{STL}{StereoLithography}
\acro{SWOT}{Strengths, Weaknesses, Opportunities and Threats}
\acro{TDC}{Time to Digital Converter}
\acro{TP}{True Positive}
\acro{TPS}{treatment planning system}
\acro{UFH}{University of Florida Health Proton Therapy Institute}
\acro{USA}{United States of America}
\acro{UPTD}{Universit\"ats Protonen Therapie Dresden}
\acro{VME}{VERSAmodule Eurocard}
\acro{WET}{Water Equivalent Thickness}
\end{acronym}

\newcommand{\API}{\ac{API}\@\xspace}

\newcommand{\CT}{\ac{CT}\@\xspace}

\newcommand{\DSC}{\ac{DSC}\@\xspace}

\newcommand{\FP}{\ac{FP}\@\xspace}

\newcommand{\GUI}{\ac{GUI}\@\xspace}

\newcommand{\LINAC}{\ac{LINAC}\@\xspace}

\newcommand{\MGH}{\ac{MGH}\@\xspace}

\newcommand{\MRI}{\ac{MRI}\@\xspace}
\newcommand{\NDA}{\ac{NDA}\@\xspace}

\newcommand{\OAR}{\ac{OAR}\@\xspace}

\newcommand{\ROC}{\ac{ROC}\@\xspace}
\newcommand{\ROI}{\ac{ROI}\@\xspace}

\newcommand{\STL}{\ac{STL}\@\xspace}

\newcommand{\TP}{\ac{TP}\@\xspace}
\newcommand{\TPS}{\ac{TPS}\@\xspace}

\begin{document}

\paper[Open-source platform for collision prevention]{An open-source platform for interactive collision prevention in photon and particle beam therapy treatment planning}

\author{F. Hueso-Gonz\'alez, P.~Wohlfahrt, D.~Craft and K. Remillard}

\address{Department of Radiation Oncology, Massachusetts General Hospital and Harvard Medical School, Boston, MA 02114, United States of America.}
\ead{fernando.hueso@uv.es}
\vspace{10pt}
\begin{indented}
\item[]March 2020
\end{indented}
\iftoggle{forarxiv}{
\vspace{10pt}
\begin{indented}
\tiny
\item[]This is an author-created, un-copyedited version of an article published in \BPEX. IOP Publishing Ltd is not responsible for any errors or omissions in this version of the manuscript or any version derived from it. The Version of Record is available online at \href{https://doi.org/10.1088/2057-1976/aba442}{https://doi.org/10.1088/2057-1976/aba442}.
\item[]\today
\end{indented}
}{}

\begin{abstract}
We present an open-source platform to aid medical dosimetrists in preventing collisions between gantry head and patient or couch during photon or particle beam therapy treatment planning. This generic framework uses the native scripting interface of the particular planning software to import STL files of the treatment machine elements. These are visualized in 3D together with the contoured or scanned patient surface. A graphical dialog with sliders allows the interactive rotation of the gantry and couch, with real-time feedback. To prevent a future replanning, treatment planners can assess in advance and exclude beam angles resulting in a potential risk of collision. The software platform is publicly available on GitHub and has been validated for RayStation with actual patient plans.
Furthermore, the incorporation of the complete patient geometry was tested with a 3D surface scan of a full-body phantom performed with a handheld smartphone.
With this study, we aim at minimizing the risk of replanning due to collisions and thus of treatment delays and unscheduled consumption of manpower. The clinical workflow can be streamlined at no cost already at the treatment planning stage. By ensuring a real-time verification of the plan feasibility, the script might boost the use of optimal couch angles that a planner might shy away from otherwise.
\end{abstract} 
\vspace{2pc}

\submitto{\BPEX}

The authors have no relevant conflicts of interest to disclose.

\maketitle

\section{Introduction}\label{sec:intro}

Treatment of cancer patients with accelerated charged particles or photon beams is performed ideally from various incidence angles \cite{bortfeld_optimization} to better spare normal tissue or \OAR surrounding the tumor. To enable the irradiation from any direction out of a $4\pi$ sphere, the treatment head is mounted on a rotating gantry, whereas the patient couch can rotate around a vertical axis, in addition to \ac{3D} translations. In the case of particle beam therapy, the treatment head (also known as nozzle) may comprise a moving snout that supports apertures, compensators and range shifters, that are positioned close to the patient surface.

As a consequence of the dynamically moving gantry head, snout and couch, there is a risk of damage of equipment, treatment interruption or even patient injury. To ensure the overall safety, aside from emergency buttons, surveillance cameras \cite{nguyen_cameras} and touch guard fins \cite{tsai_stereotactic}, potential collisions between gantry head and couch or patient need to be assessed in advance and prevented \cite{hoopes_incidents}. Throughout the last three decades, different approaches have been developed to aid treatment planners in the avoidance of irradiation angles with risk of collision. These were based on simplified analytical calculations \cite{yorke_collision,humm_collision,muthuswamy_collision,beange_collision,furhang_clearance,niotsikou_collision,hua_collision}, graphical simulations \cite{sherouse_simulator,kessler_collision,bayouth_collision,tsiakalos_collision,ward_collision,hamza_collision,hamzalup_collision,glaser_virtual,zou_collision,yu_noncoplanar,suriyakumar_slicer,maclennan_collision} or experimental reference measurements \cite{chao_avoidance,brahme_collision,becker_varian,becker_elekta,padilla_kinect,jung_proton,cardan_cameras,lang_collision}. In some cases, the combination of treatment parameters leading to a collision are depicted as keep-out areas in a set of reference charts, or implemented as a warning feedback within the treatment software. In others, the user can move the isocenter and rotate the gantry interactively, and the risk of collision is detected automatically or assessed visually.

Collision detection during treatment planning is one important tool in the context of personalized medicine, where the optimum treatment plans for every patient are sought, but must be feasible at the same time.
Despite extensive research and the abundant number of sophisticated solutions proposed by individual hospitals or vendors during the last three decades, there is no standardized solution applied in radiotherapy centers. In many cases, radiotherapy departments lack of embedded collision assistance during the treatment planning stage. For example, at \MGH, dosimetrists rely on experience and eyeball intuition as to which incidence directions and isocenter positions are infeasible. Furthermore, the therapists check for collisions during a dry run with the patient at the actual \LINAC.
The two main consequences, as stated by other authors \cite{furhang_clearance,ward_collision,hamzalup_collision,glaser_virtual,padilla_kinect,zou_collision,jung_proton,yu_noncoplanar}, are:
\begin{itemize}
\item[] \emph{Time:} The need for dry runs to ensure patient safety decreases the time for patient treatments, and requires a slight time dedication by the therapists. Furthermore, in the case of a collision and thus infeasible treatment plan, there is an unexpected delay in the treatment start, which alters notably the scheduled clinical workflow and consumes a critical amount of manpower. A planner has to devote several hours to redo the treatment plan with new beam and couch geometries, which need to be checked by medical physicists and approved by clinicians again. Finally, a new dry run is required.
\item[] \emph{Dose:} Treatment planners tend to be conservative when choosing beam angles, in order to minimize the probability of replanning due to a potential collision during dry run. This ensures a smooth clinical workflow and reliable schedule, but prevents a full exploitation of the capabilities and conformality of the patient-specific therapy. Furthermore, in the case of non-existing \ac{3D} modeling tools, the planner faces more difficulties in the visualization of the treatment and might shy away from introducing couch angles and non-coplanar irradiation \cite{yu_noncoplanar} that might be beneficial from the dosimetric perspective. 
\end{itemize}
Despite the numerous published solutions, these have only been applied scarcely in few institutions so far, and are not being used routinely in the standard clinical workflow or commercial \ac{TPS}.
Analytical solutions are fast and handy, but are not patient-specific \cite{cardan_cameras}. In cases where the patient geometry is incorporated from a \CT scan, it is incomplete and collisions might occur with body parts outside the field of view, in particular extremities. Proposed workarounds include the use of additional \ac{3D} scanners or cameras, but these are not installed (yet) by default in \CT scanners, and thus require some investment and setup efforts.
Furthermore, there might be a lack of coordination between independent vendors providing the \TPS software, the treatment machine and the patient couch for delivery. The respective information should be combined coherently in the same platform ensuring an effective collision detection workflow.
Indeed, the \ac{3D} model of the gantry might not be available or disclosed by the vendor if not requested upon purchase \cite{tsiakalos_collision}. In this case, measurements or \ac{3D} scanning of the treatment head require expertise, hardware and software integration, which increase costs and efforts \cite{cardan_cameras, danuser_cameras}. For hospitals having treatment machines from different vendors, the integration effort is multiplied.
On the other hand, some vendors include powerful \ac{3D} visualization tools for real-time interaction with the delivery machine. However, these are not usually embedded in the planning software. Also, external programs from third-party vendors \cite{yu_noncoplanar} add licensing costs. In the case of open-source alternatives like Slicer3D \cite{kikinis_slicer3d}, whose SlicerRT module \cite{pinter_slicerRT} includes collision detection \cite{suriyakumar_slicer}, its application forces data transfer and efforts by the dosimetrist if the planning is done in a commercial software.

At the Department of Radiation Oncology at \MGH, for example, RayStation (RaySearch Laboratories AB, Stockholm, Sweden) is used for radiotherapy treatment planning. It provides an embedded \ac{3D} visualization tool of the patient, with a default simplified treatment machine showing the beam incidence. Some users have privately developed basic collision detection scripts in RayStation by modeling the gantry geometry as combinations of boxes and cylinders \cite{maclennan_collision,ho_collision}.

Building upon this experience and the aforementioned pitfalls, this paper proposes \emph{RadCollision}, an open-source platform for collision assessment in \TPS that:
\begin{itemize}
\item Is licensed under GPLv3 \cite{gplv3} at no cost, and can be downloaded online,
\item Is maintained by the scientific and clinical community through public repositories,
\item Can progressively support \TPS from further vendors,
\item Is easily adaptable by any institution,
\item Does not require purchasing additional hardware,
\item Does not need expert knowledge about software,
\item Is embedded in every \TPS and does not require external software or data transfer to other servers,
\item Is patient-specific,
\item Provides a realistic \ac{3D} visualization of nozzle, couch and patient, rather than reference charts,
\item Is modular, so that further room elements can be added into (or removed from) the visualization by the end user,
\item Depends on \ac{3D} model input files in \STL format,
\item Aids treatment planners in choosing beam angles with interactive sliders,
\item Allows the independent movement of each treatment room element, with real-time feedback,
\item Prioritizes speed over precision and sophistication in order to boost its integration in the clinical workflow,
\item Offers the choice between automatic or visual collision detection, the latter relying on the planner's ability to assess the collision risk in incomplete patient geometries,
\item Relies on an initial \ac{3D} modeling of the treatment machine, or the willingness of vendors to provide their \ac{3D} models to hospitals,
\item Optionally incorporates the full patient geometry recorded with any \ac{3D} scanner or surface imaging device.
\end{itemize}

This manuscript is organized as follows. The software framework and details of its implementation are discussed in section~\ref{sec:methods}. The application of the platform for the RayStation scripting interface is validated in section~\ref{sec:results}. A brief discussion and the main conclusions of the paper are presented in sections~\ref{sec:discussion} and \ref{sec:conclusion}.
 
\section{Materials and Methods}\label{sec:methods}

\subsection{Software architecture and system description}\label{ss:architecture}

\begin{figure}[ht]
\centering

\begin{tikzpicture}[node distance = 1.25cm]

\node [save path=\tmpath, block,rectangle split part fill={orange!20,gray!5},font=\fontsize{10}{0}\selectfont, text width=4.3cm,
label={[,font=\fontsize{9}{0}\selectfont,anchor=south]above:Room-specific}] (MachineModel) {\textbf{Machine model}
\nodepart{two}
\vspace{-2mm}
\begin{itemize}[leftmargin=4mm]
\item \mystrut Vendor files under NDA
\item \mystrut Online \ac{3D} repositories
\item \mystrut Measurements, \ac{3D} scans
\item \mystrut Cylinders and boxes
\end{itemize}};

\begin{scope}[on background layer]  
\fill[black!50,opacity=0.5,
use path=\tmpath,transform canvas={xshift=0.5ex,yshift=-0.5ex}];
\end{scope}
\begin{scope}[on background layer]  
\fill[black!50,opacity=0.5,
use path=\tmpath,transform canvas={xshift=1ex,yshift=-1ex}];
\end{scope}

\node [block,rectangle split part fill={orange!20,gray!5},font=\fontsize{10}{0}\selectfont, text width=4.1cm, right of=MachineModel, xshift=3.7cm,
label={[,font=\fontsize{9}{0}\selectfont,anchor=south]above:Patient-specific}] (PatientModel) {\textbf{Patient model}
\nodepart{two}
\vspace{-2mm}
\begin{itemize}[leftmargin=4mm]
\item \mystrut \CT or \MRI scan
\item \mystrut Phantom templates
\item \mystrut Fixed \ac{3D} cameras
\item \mystrut Smartphone scan
\end{itemize}};

\path (MachineModel) -- (PatientModel) node[midway,yshift=-1.9cm,font=\fontsize{9}{0}\selectfont] (STLJoint) {Conversion};

\draw [->,thick,>=latex] (MachineModel) -- (STLJoint);
\draw [->,thick,>=latex] (PatientModel) -- (STLJoint);

\node [fill=green!20,draw,font=\fontsize{10}{0}\selectfont, below of=STLJoint, yshift=7mm] (STL) {\textbf{\STL format}};

\node [block,rectangle split part fill={orange!20,white},font=\fontsize{10}{0}\selectfont, text width=4.3cm, below of=PatientModel, yshift=-6.92cm,xshift=-8pt] (Interface) {\textbf{Scripting interface}
\nodepart{two}
\vspace{-2mm}
\begin{itemize}[leftmargin=4mm]
\item \mystrut Import \STL as \ac{ROI}
\item \mystrut Transform \ac{ROI} $4\times4$
\item \mystrut Calculate \ac{ROI} overlaps
\item \mystrut Multi-thread support
\end{itemize}};

\node [whtblock, below=of Interface,font=\fontsize{10}{0}\selectfont, yshift=8mm] (viewer) {\ac{3D} visualization of \ac{ROI}s};

\begin{scope}[on background layer]
  \coordinate (aux1) at ([yshift=17mm]Interface);
  \node [container,fit=(aux1)(viewer)(Interface)] (TPS) {};
  \node at (TPS.north) [fill=white,draw,font=\fontsize{12}{0}\selectfont] {\textbf{\TPS}};
\end{scope}

\node [block,rectangle split part fill={orange!20,white},font=\fontsize{10}{0}\selectfont, text width=4.3cm, left of=MachineModel, xshift=-4cm, yshift=-0.8cm,
label={[,font=\fontsize{9}{0}\selectfont,anchor=south]above:Generic}] (CoreLayer) {\textbf{Core layer}
\nodepart{two}
\vspace{-2mm}
\begin{itemize}[leftmargin=4mm]
\item \mystrut Abstract classes \lstrut and methods \lstrut
\item \mystrut Rotation matrices \lstrut and translation vectors \lstrut
\end{itemize}};

\node [save path=\tmpath, block,rectangle split part fill={orange!20,white},font=\fontsize{10}{0}\selectfont, text width=4.3cm, below of=CoreLayer, yshift=-2.5cm,
label={[,font=\fontsize{9}{0}\selectfont,anchor=south]above:\TPS-specific}] (InterfaceLayer) {\textbf{Interface layer}
\nodepart{two}
\vspace{-2mm}
\begin{itemize}[leftmargin=4mm]
\item \mystrut \TPS-specific \lstrut \API calls instantiating \lstrut core\\methods \lstrut
\item \mystrut \GUI dialog \lstrut creation: object \lstrut selection,\\ movement \lstrut sliders and collision \lstrut report
\end{itemize}};

\node [whtblock, below=of InterfaceLayer,font=\fontsize{10}{0}\selectfont, yshift=6mm, minimum height=4.5em] (Startup) {\emph{Startup sequence:} \lstrut ask to select \lstrut active room elements \lstrut and load \ac{3D} models \lstrut as \acp{ROI}};

\node [whtblock, below=of Startup,font=\fontsize{10}{0}\selectfont, yshift=11mm, minimum height=4.5em] (Execution) {\emph{Execution loop:} \lstrut upon change \lstrut in \GUI dialog settings, transform \lstrut \acp{ROI} and recalculate \lstrut collision};

\node [block,rectangle split part fill={orange!20,white},font=\fontsize{10}{0}\selectfont, text width=4.3cm, below of=MachineModel, yshift=-3.3cm,
label={[,font=\fontsize{9}{0}\selectfont,anchor=south]above:Hospital-specific}] (SetupLayer) {\textbf{Setup layer}
\nodepart{two}
\vspace{-2mm}
\begin{itemize}[leftmargin=4mm]
\item \mystrut Definition \lstrut of available treatment \lstrut machines
\item \mystrut Storage \lstrut directory of \STL files \lstrut
\end{itemize}};

\begin{scope}[on background layer]
  \coordinate (aux2) at ([yshift=7mm]CoreLayer.north);
  \node [container,fill=blue!15,fit=(aux2)(CoreLayer)(InterfaceLayer)(Startup)(Execution)] (RadCollision) {};
  \node at (RadCollision.north) [fill=white,draw,font=\fontsize{12}{0}\selectfont] {\textbf{Open-source platform}};
\end{scope}

\begin{scope}[on background layer]  
\fill[black!50,opacity=0.5,
use path=\tmpath,transform canvas={xshift=0.5ex,yshift=-0.5ex}];
\end{scope}
\begin{scope}[on background layer]  
\fill[black!50,opacity=0.5,
use path=\tmpath,transform canvas={xshift=1ex,yshift=-1ex}];
\end{scope}

\draw [<->,>=latex,double,thick] (CoreLayer.188) to [out=-135,in=135] (InterfaceLayer.162);
\draw [<->,>=latex,double,thick] (InterfaceLayer.214) to [out=-135,in=135] (Execution.160);
\draw [<->,>=latex,double,thick] (Startup.10) to [out=45,in=-45] (InterfaceLayer.-17);
\draw[->,thick,>=latex,rounded corners=5pt] (SetupLayer.275)|-(Interface.170)  ;   
\draw[<-,thick,>=latex,rounded corners=5pt] (SetupLayer.south)|-($(Startup.3) + (0mm,0.1mm)$)  ;
\draw[->,thick,>=latex,rounded corners=5pt,double] (STL.330)|-(SetupLayer.345)  ;
\draw[->,thick,>=latex,rounded corners=5pt] (Execution.east)--($ (Interface.182) + (-5mm,-1mm) $)  ;   
\draw[->,thick,>=latex] (Interface.182) to [out=-180,in=180,xshift=-6mm] (Interface.194);

\end{tikzpicture} \caption{Proposed software architecture \emph{RadCollision} for collision prevention during treatment planning.}
\label{fig:architecture}
\end{figure}
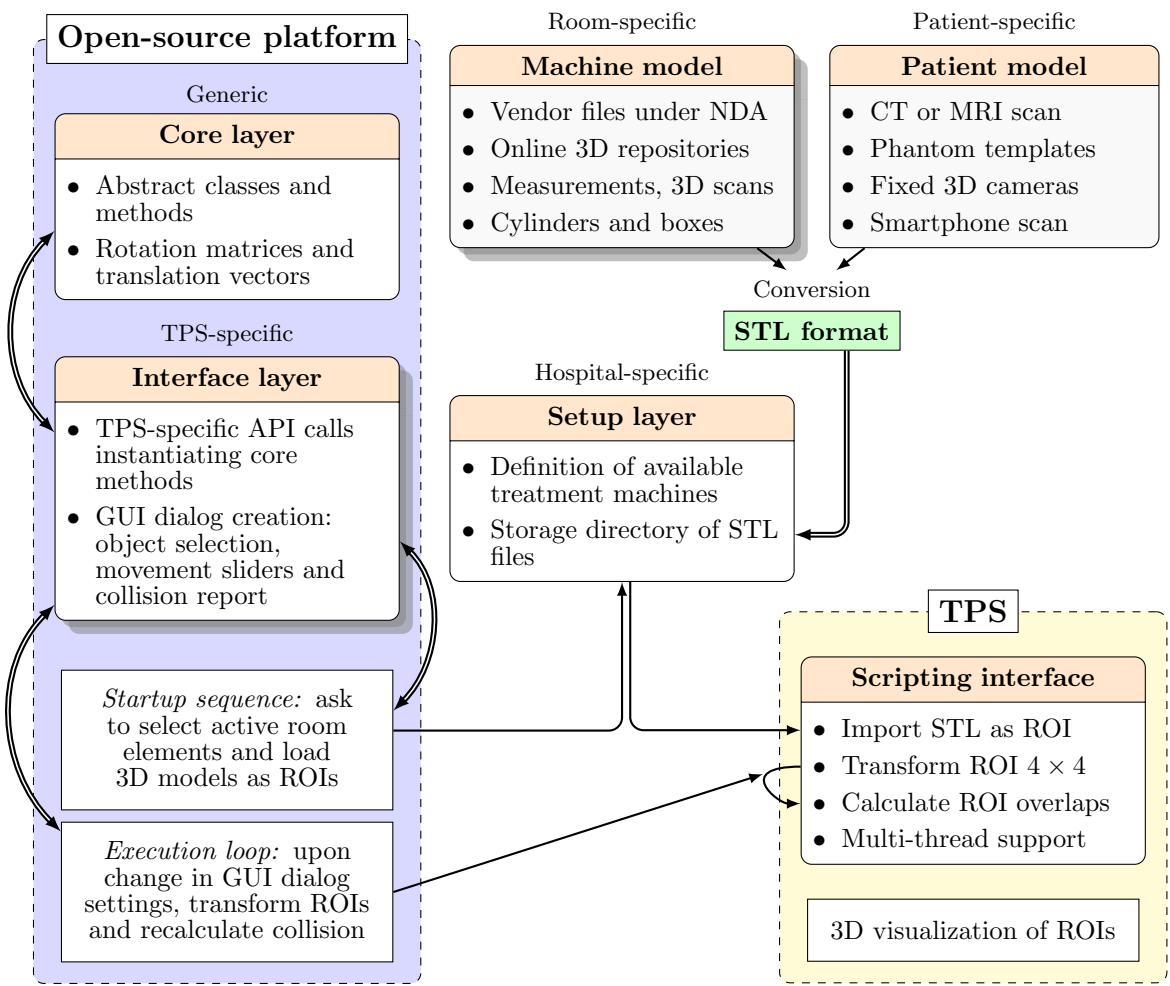

The proposed software model for collision prevention is illustrated in \refig{fig:architecture}. It is designed to be as embedded as possible into the \TPS used by the dosimetrist, but keeping the flexibility and modularity, as specified in section~\ref{sec:intro}.

First, it is assumed that the \ac{3D} outer surfaces of each treatment machine and any other room elements relevant for collision are available to the hospital. These may be requested to the vendors upon purchase or acquired later under an \NDA. Or they can be downloaded from online \ac{3D} stores or community repositories. Alternatively, one may generate these in situ based on a \ac{3D} scan of the machine, or experimental measurements \cite{tsiakalos_collision,maclennan_collision,ho_collision}.
Regardless of the source, the 3D model shall be centered at the room isocenter, processed to remove unnecessary internal sub-parts, and exported as \STL format \cite{stl_format}, one for each subpart of the machine moving independently. These files are stored in a shared directory of the hospital servers.

Second, the model of the patient relies on the external contour of the \CT or \MRI dataset of the patient, which is usually already available as a \ROI within the \TPS and thus no specific action is needed. If a more complete model of the patient is required, phantom-based extensions \cite{niotsikou_collision}, in-room \ac{3D} cameras \cite{cardan_cameras} or even scans from handheld devices, see subsection~\ref{ss:scan3d}, may be used. These \ac{3D} scans need to be converted to \STL format with \eg the Meshlab open-source software \cite{cignoni_meshlab} and imported into the \TPS as external contour.

Third, it is required that the deployed \TPS software comprises an embedded viewer of \acp{ROI} as \ac{3D} surfaces and provides a scripting interface with multi-thread support. Three public methods are essential to support this application: the ability to import an \STL file as an \ROI, to transform (rotate and translate) any \ROI with a $4\times4$ matrix, and to calculate the region of overlap between two \acp{ROI}.

Considering this set of prerequisites, we propose and design a new open-source software online platform, named \emph{RadCollision}, as a generic tool for collision prevention in radiotherapy. It is divided in a core layer and an interface layer, whereas the setup layer lies outside of the public platform.

Its core layer defines the abstract classes and methods. For example, an element rotating around isocenter, like the gantry head, or any object translating in \ac{3D} and rotating around the vertical axis, like the couch. This layer also generates the corresponding transformation matrices depending on the irradiation angle or couch position according to the DICOM (IEC 61217) coordinate system conventions \cite{dicom_standard}.

The interface layer handles the \GUI, as well as the communication with the \API functions of the specific \TPS. Because the function signatures might differ, and each \TPS may support a different programming language within their scripting interface, this layer may have to be duplicated and specialized for every case (see shadow in \refig{fig:architecture}), wrapping the calls to the generic core methods.

The setup layer is hospital-specific and consists of a database of all \ac{3D} models of the available machines and any other relevant room elements. Each part has to be assigned to one of the abstract classes defined in the core layer according to its particular motion behavior (degrees of freedom).

During treatment planning, once the patient is contoured, the user can start the collision prevention software. The program automatically chooses the machine and couch model from the active treatment plan among those available in the hospital database (setup layer). The selected ones are loaded as \acp{ROI} by the \TPS (scripting interface). Then, the \GUI dialog (interface layer) allows for the adjustment of irradiation settings (gantry angle, couch angle, snout extraction, etc.). The software transforms in real-time the \acp{ROI} corresponding to the treatment machine, and calculates any collision (overlap of \acp{ROI}) with the patient or couch in the background. There is also a \GUI button to automatically calculate the collision report for each beam defined in the treatment plan. For dynamic arcs, the collision is calculated in steps of one degree.

\subsection{Implementation for RayStation}\label{ss:implementation}

We exemplarily validated the proposed software model for the \TPS RayStation. The interface layer was written in IronPython \cite{foord_ironpython}, the original implementation language of the scripting library of RayStation. The threaded \GUI relies on the native WinForms library \cite{sells_winforms}. The script is publicly available on the \MGH radiation oncology GitHub organization \cite{mghro_github}, and requires the use of RayStation 8B or newer versions.

The main three functions from the RayStation \API called by the interface layer are:
\begin{enumerate}
\small
\item \code{ImportRoiGeometryFromSTL(FileName,TransformationMatrix)}
\item \code{TransformROI3D(TransformationMatrix)}
\item \code{ComparisonOfRoiGeometries(RoiA,RoiB,ComputeDistanceToAgreementMeasures)}
\end{enumerate}

The import of the \STL file (function 1) is done only once, at script startup. This function is available since RayStation version 8B. Each time a slider of the \GUI is changed, the \ac{3D} transformation (function 2) has to be applied on the already imported \ROI. This $4\times4$ affine transform matrix, specifically defined for the treatment isocenter, is computed independently for each sub-part of the couch or nozzle according to the motion behavior initially configured by the user (setup layer). If automatic collision detection is enabled in the \GUI, the third function calculates if two \acp{ROI} overlap via the \DSC \cite{dice_dsc}.

It shall be noted that, as the \ac{3D} modeling in RayStation is done in the patient coordinate system, the simulation of couch angles is done by rotating the room elements (gantry and optionally walls) in the opposite direction rather than by rotating the couch model.

\subsection{Optional 3D surface scan}\label{ss:scan3d}
The presented framework, \cf \refig{fig:architecture}, is compatible with the import of a \ac{3D} surface scan of the full patient geometry, in order to detect potential collisions with parts of the body outside the field of view of the \CT scan. Except for the requirement to export the \ac{3D} surface scan as \STL file, no prior assumptions are needed for the scanning device. Finally, the \ac{3D} scans have to be rigidly registered to the \CT scan geometry.

To illustrate this workflow, we acquired a \CT scan of an anthropomorphic female phantom (Alderson Research Laboratories, Stanford, CT, USA) using a GE Discovery RT \CT scanner (GE Healthcare, Chicago, IL, USA), which served as ground-truth geometry. Subsequently, a \ac{3D} surface scan with a handheld iPhone XS (Apple, Cupertino, CA, USA) was performed. The front face camera of the smartphone comprises depth sensor technology \cite{apple_truedepth}, that can be used in combination with the free application \emph{Capture: 3D Scan Anything} (Standard Cyborg, Inc, San Francisco, CA, USA) to obtain a \ac{3D} surface scan of an object in \ac{PLY} format. The \ac{PLY} file can be imported into \eg Slicer3D for rigid registration with the \CT scan geometry, and the resulting mesh can be exported as \STL file (or even directly as contour in an RT structure file). A similar procedure can be conducted with any other \ac{3D} scanner type.

The conformity of external contours derived from the \CT scan and the \ac{3D} surface scan was assessed by the minimal contour displacement and Hausdorff distance, defined as the 95th quantile of absolute contour distances for each axial \CT slice from head to pelvis of the anthropomorphic female phantom \cite{taha_metrics}. 
 
\section{Results}\label{sec:results}

\begin{figure}[ht]
\begin{center}
\includegraphics[width=\textwidth]{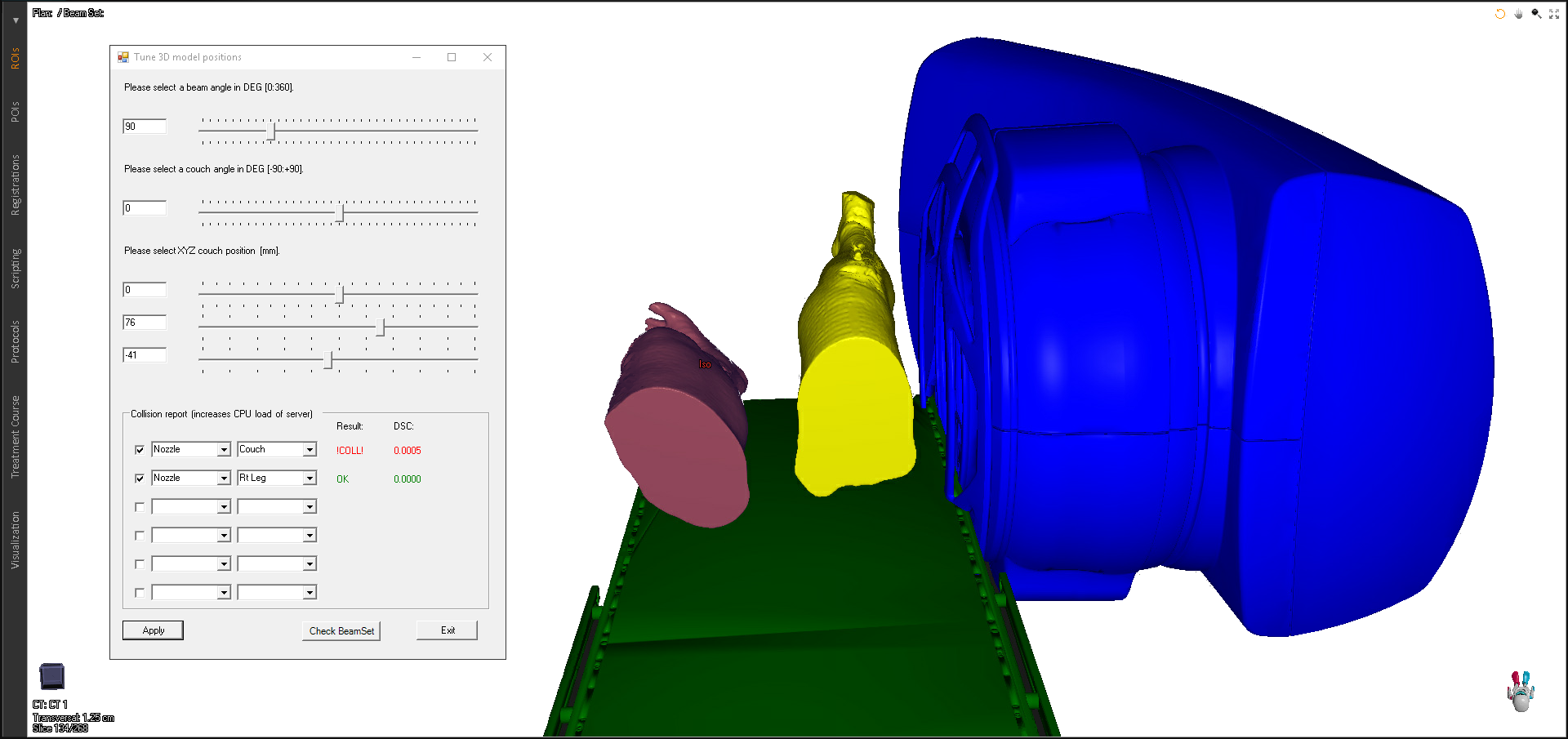}
\caption{Illustration of a collision between gantry (blue) and couch (green) at a gantry angle of 90 degrees detected in the \ac{3D} viewer tab of RayStation. The external patient contours are visible together with the imported \STL files of the \LINAC and couch of the Elekta Agility radiotherapy treatment room. The \ac{3D} models of the machine parts provided by Elekta (Stockholm, Sweden) had been manually preprocessed and converted to \STL. The \GUI dialog of the running script allows for a real-time adjustment of the couch position, couch angle and gantry angle interactively via five independent sliders. The axis of rotation crosses the treatment isocenter defined by the planner. A collision report warns about a collision of gantry with couch, whereas none is found with the right leg (yellow contour).\label{fig:collision}}
\end{center}
\end{figure}

The implementation of \emph{RadCollision} for the RayStation \TPS was evaluated qualitatively with actual patient plans from the \MGH radiotherapy department.
Four patient plans, which were found infeasible during collision check by the therapists in the past, were analyzed retrospectively. Based on the \ac{3D} visualization and the collision report results, \cf \refig{fig:collision}, collisions with the couch were found at similar angles than those reported experimentally (within 2 degrees). The replanned treatments (with other beam angles or isocenter positions) were also studied, showing no effective collision for the selected incidence directions. In a fifth case, the simulation was applied prospectively, before patient treatment, and the predicted absence of collisions was confirmed during a dry run with the patient in position. A quantitative analysis of the overall prediction accuracy was not performed, as it depends on the input data (more details in section~\ref{sec:discussion}) rather than on the proposed software model. Hence, the latter has been the focus of this manuscript.

\begin{figure}
\begin{center}
\includegraphics[width=\textwidth]{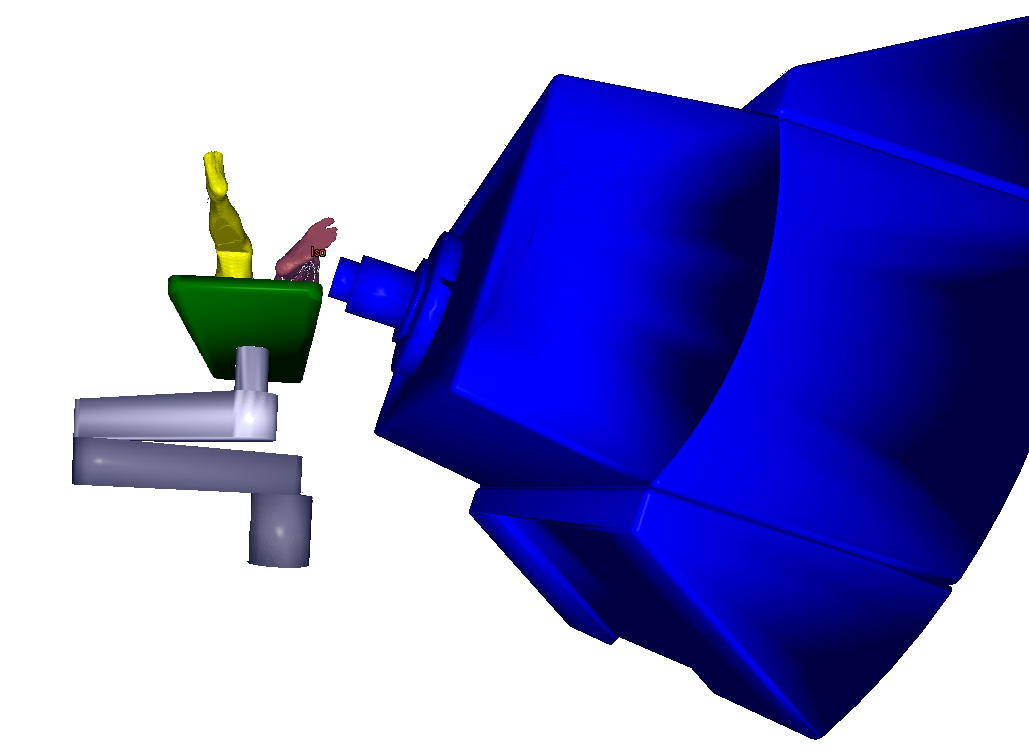}
\caption{Assessment of collisions between proton therapy nozzle (blue) and patient couch (green) or positioning robot (gray) in the \ac{3D} viewer tab of RayStation. The position of the articulated robot arms is recalculated in real-time whenever the couch position is changed via the \GUI dialog of the running script, \cf \refig{fig:collision}. The \ac{3D} models were manually created based on experimental measurements at a proton therapy gantry treatment room. \label{fig:proton}}
\end{center}
\end{figure}

In \refig{fig:proton}, the software was tested with the model of a proton treatment room and a robotic system for patient positioning consisting of two articulated arms. The robot configuration was automatically calculated based on the couch position, in order to assess the collision risk between the robot arms and the nozzle.
The interactivity capabilities of the \GUI are shown in \refig{fig:videos} for photon therapy (top) and proton therapy (bottom). 

\begin{figure}
\includegraphics[width=\textwidth]{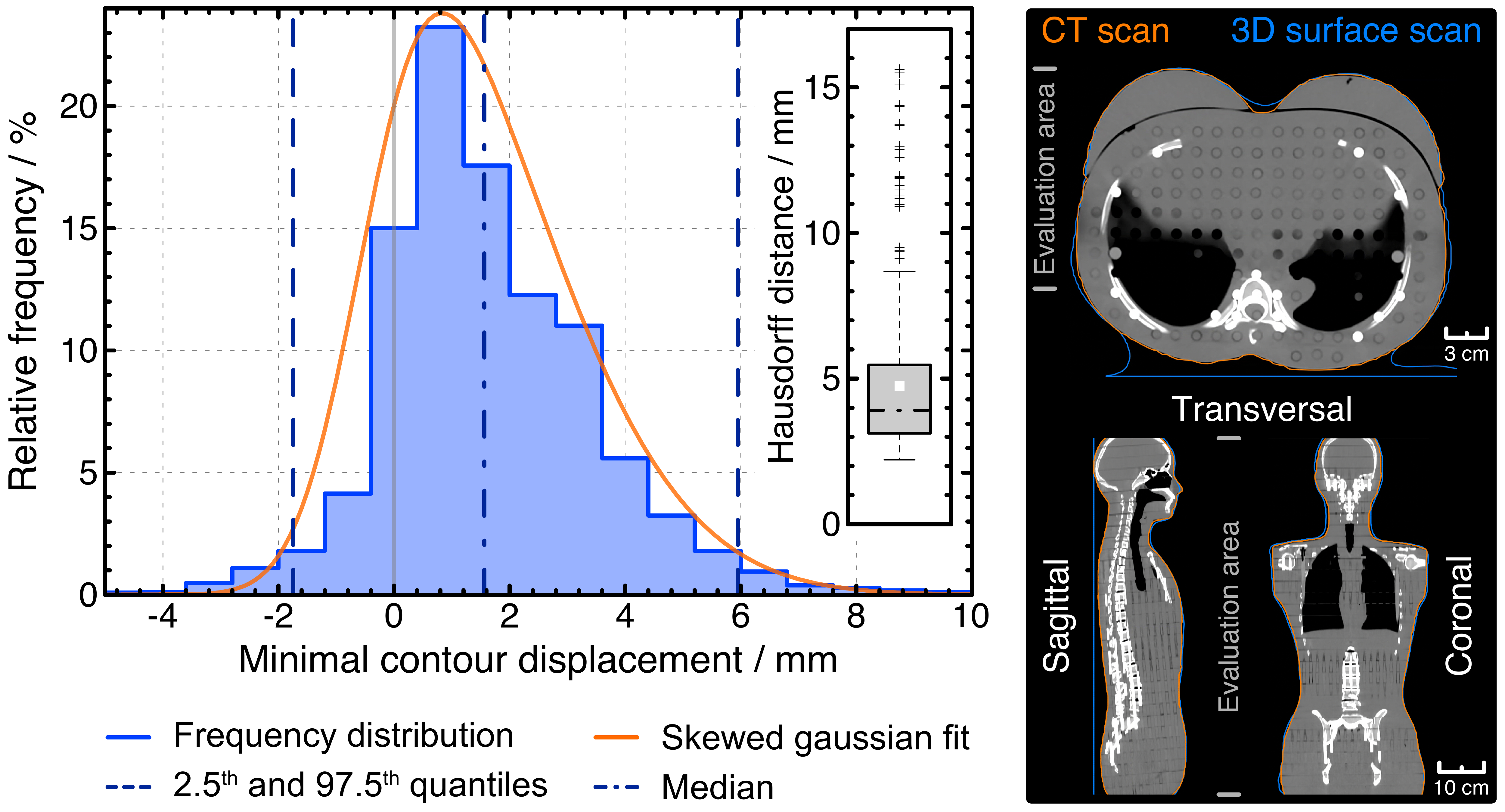}
\caption{Quantitative assessment of the minimal displacement of the external contour derived from a \CT scan (ground-truth) and a \ac{3D} surface scan (\refig{fig:scan3d}). The distribution of the Hausdorff distance (95th quantile) determined in the evaluation area of each axial \CT slice from head to pelvis is summarized as box plot. \label{fig:hausdorff}}
\end{figure}

The quantitative analysis of the accuracy of the \ac{3D} surface scan (\refig{fig:scan3d}) with respect to the ground-truth geometry derived from a \CT scan is shown in \refig{fig:hausdorff}. The geometry obtained by the \ac{3D} surface scan is in general slightly larger than the \CT geometry, which provides more conservative results for the collision test. The median distance between the two external contours in the evaluation area, excluding the region of contact between patient and couch surface, is roughly 1.6\,mm. In 8\% of all cases, the contour pixels from the \ac{3D} surface scan are inside of the external contour determined on the \CT scan (negative minimal contour displacement) with a mean absolute deviation of $(-1.6 \pm 1.0)$\,mm. Overall, the minimal contour displacement was within -1.7\,mm (2.5th quantile) and 5.9\,mm (97.5th quantile) at a 95\% confidence level. Differences larger than 10\,mm were occasionally observed for some \CT slices, in particular in the neck region, which were mainly caused by a non-optimal orientation of the smartphone during the \ac{3D} surface scanning test.

\section{Discussion}\label{sec:discussion}

This manuscript proposes \emph{RadCollision}, a potential generic solution for collision assessment in a variety of treatment modalities by importing \STL files of the machine and room elements through the scripting interface of a \TPS, \cf \refig{fig:architecture}. This approach is as embedded as possible in the workflow of dosimetrists, open-source, modular, and does not imply any investment for the hospital.

However, this framework requires some coordinated initial efforts from several parties. First, the different \TPS softwares have to support the import of \STL files as \ROI through their scripting interface, \cf \refig{fig:architecture}, as well as to enable a \ac{3D} viewer tool. To date, we are only aware of Slicer3D and RayStation \acp{TPS} to offer these functionalities.
Second, to obtain the highest precision, the treatment machine vendors have to provide the \ac{3D} models to the hospitals under an \NDA.
Third, hospital staff has to process and organize these models into a database (see setup layer).
Fourth, the scientific open-source community has to write a specific variation of the software for each \TPS vendor (see interface layer).
Once all of these prerequisites are fulfilled and the initial setup is performed, the treatment planner will be able to routinely deploy an embedded collision detection tool within their normal workflow with no effort.

Other collision detection methods published in the literature are more specific and sophisticated than the presented solution, but also more complex to implement and require the acquisition of further hardware like fixed cameras or room lasers,
and potentially the use of external proprietary software \cite{yu_noncoplanar} and the need of data transfer from the \TPS. This might be an obstacle for implementation in a widespread context.
In contrast, the \emph{RadCollision} framework is embedded (provided a set of prerequisites), modular and scalable, by allowing through the setup layer (\refig{fig:architecture}) the progressive addition of other sub-elements of the treatment room like electron applicators or imaging detector panels \cite{davis_collision}, without the need of upgrading the \TPS software.
It also allows for (but does not force to) the incorporation of the complete \ac{3D} patient surface, and is agnostic about the 3D scanning source, \eg a handheld smartphone (\refig{fig:scan3d}), as long as the output is converted to \STL format.

It should be noted that the reliability of the collision assessment within this software platform depends on the accuracy of the input data, \cf \refig{fig:architecture}, rather than on the software itself (numeric rounding aside). The following sources of uncertainty are identified:
\begin{itemize}
\item \ac{3D} models of the machine and couch
\item \CT scan of the patient (pixel size, image quality, artifacts, contouring)
\item \ac{3D} scan (if performed)
\item Patient positioning (treatment versus imaging)
\item Patient anatomy (variations over the treatment course)
\item Patient motion  (variations within a treatment fraction)
\end{itemize}

In general, \ac{3D} models of the machine elements provided by the vendors are very precise (manufacturing tolerance and specifications), whereas the patient representation has a higher error, either due to the restricted field of view of the \CT scan, or due to the inaccuracy of the \ac{3D} scan of the patient surface, \cf \refig{fig:hausdorff}.

To use this tool for patients, reasonable safety margins for collisions should be introduced by inflating the external \ROI in the \TPS. The overall collision prediction accuracy, \ie \TP rate, will be a combination of the aforementioned uncertainties and the chosen safety margin. The specific choice should be in accordance with the estimated magnitude of these errors, that might be specific for each machine, \ac{3D} scanner, patient age (motion) and tumor site.
In a clinical setting, a way to calibrate the safety margin could be to draw the \ROC curve of the collision prediction in an initial study for different margins and cases, and find a compromise between maximizing the sensitivity and minimizing the \FP rate.

This generic software paradigm was realized for the \TPS RayStation, using IronPython as scripting language. A \GUI dialog with sliders (\refig{fig:collision}) allows for an interactive adjustment of beam angle, couch angle and snout extraction with real-time feedback, and reports the risk of collision. The automatic collision detection runs on a separate thread pool, not to freeze the feedback of the \GUI, \cf \refig{fig:videos}. Nonetheless, it can be switched off by the user for reducing the overall server load \cite{zou_collision}, if needed. A \GUI button triggers the calculation of the collision report for all the beams and arcs in the treatment plan.
The code is openly available in a GitHub repository \cite{mghro_github} and can be maintained by the collaborative efforts of the scientific and clinical communities.

The integration of this tool in the clinical routine of a radiotherapy department might contribute to an overall improvement of the daily workflow: less or no time is required for actual collision checks using the treatment machine, thus reducing the workload of therapists as well as the machine time not available for patient treatments. Moreover, delays in the beginning of patient therapy are prevented, which otherwise emerge when a collision is found during the dry run at the treatment room, requiring unscheduled allocations of time and resources for replanning.

Furthermore, the \ac{3D} visualization of the actual treatment room at the planning stage facilitates the selection of optimal beam and couch angles, which can in turn improve the dosimetric quality of the plan. By providing a real-time assurance that the selected angles do not present a risk of collision, \ie a risk of cost-intensive replanning, the dosimetrists are less likely to shy away from irradiation geometries beneficial from the dose perspective. In this regard, the script could be most helpful for clinical cases such as stereotactic treatments, extremities, partial breast irradiation and prone breast treatments, electron beams, as well as plans with drastically anterior or posterior isocenters. The presented tool is expected to aid in the development of optimally individualized treatment plans.

In the future, users of other \TPS software might contribute to the public repository writing their specific interface layer, \cf \refig{fig:architecture}, and request their vendors to support this software model in their future releases. Namely, a \ac{3D} viewer tab and three specific functions would need to be implemented on their side: the ability to import a \ac{3D} model \STL file as an \ROI, the transformation of \acp{ROI} based on an affine transform matrix, and the calculation of the overlap between two contours.

\section{Conclusions}\label{sec:conclusion}

An open-source software architecture for patient-specific collision assessment in external beam radiotherapy is proposed. It relies on the native scripting interface of each \TPS, and requires its ability to import \STL files of the patient couch and treatment head as \acp{ROI}. These are superimposed with the contoured patient geometry in the \ac{3D} visualization tab.
It also enables the incorporation of the complete patient geometry, that might not be fully represented in the underlying \CT scan, based on any \ac{3D} surface scanning device. This can aid the planner in estimating whether the treatment head will collide with any part of the patient, for example with the arms of breast patients. Hence, it minimizes the risk of replanning and thus of treatment delays, and allows for the choice of optimum and feasible irradiation angles.

The presented collision detection tool was evaluated for the RayStation \TPS with actual patient plans, with no additional external software required. It will be included as part of the clinical workflow of dosimetrists of the radiotherapy section of \MGH as soon as an upgrade to RayStation version 8B (or higher) for clinical use is performed. Future work will be devoted to the inclusion of the \ac{3D} models of accessories like wedges and boli, and to the automatic feasibility assessment of final beam angles and dynamic arcs as a prerequisite for the treatment plan approval. 
\vfill
\ack{
We thank T.~Bortfeld, D.~M.~Edmunds, D.~Gierga, D.~Kunath, A.~Lasso, R.~L\"oschner, M.~Luzzara, G.~Sharp, J.~Smeets, J.~S\"oderberg, M.~Spiegel, K.~St\"utzer, J.~M. Verburg, E.~Vidholm, S.~Yan and W.~Zou for scientific advice and discussions, and the Lunder team at MGH for technical support. This work was supported in part by the Federal Share of program income earned by Massachusetts General Hospital on C06-CA059267, Proton Therapy Research and Treatment Center.
} 
\clearpage
\iftoggle{estiloharvard}{}{} 
\clearpage

\renewcommand{\thesection}{S}
\renewcommand{\thesubsection}{S\Alph{subsection}} 
\renewcommand{\leftmark}{SUPPLEMENTARY MATERIAL}

\newcommand{\ResetCounters}{
\setcounter{table}{0}
\setcounter{figure}{0}
\renewcommand{\thefigure}{S\arabic{figure}}
\renewcommand{\theHfigure}{Supplement.\thefigure}
}

\section{Supplementary Material}
\ResetCounters

\begin{figure}[hb]
\begin{center}
\includemedia[
     width=\textwidth,
     height=0.5625\textwidth,
     addresource=linac_v2f.mp4,
     flashvars={
         source=linac_v2f.mp4
     }
]{}{VPlayer.swf}

\vspace{1cm}

\includemedia[
     width=\textwidth,
     height=0.5625\textwidth,
     addresource=proton_v2f.mp4,
     flashvars={
         source=proton_v2f.mp4
     }
]{}{VPlayer.swf}
\caption{Video examples of the collision assessment in photon beam radiotherapy (top) or proton therapy (bottom) based on the presented open-source script in the \ac{3D} viewer tab of RayStation. Note: the figure may appear blank unless opened with Adobe PDF Reader with \url{https://get.adobe.com/flashplayer/} installed.\label{fig:videos}}
\end{center}
\end{figure}

\begin{figure}
\begin{center}
\includemedia[
  label=Alderson 
  ,width=\linewidth
  ,height=1.2\linewidth
  ,3Dtoolbar
  ,3Dmenu
  ,3Dcoo=-0.026843339204788208 -0.12900663912296295 0.6215205788612366
  ,3Dc2c=-0.3824044167995453 0.24747537076473236 -0.89023756980896
  ,3Droo=1.511344133826188
  ,3Droll=-163.9625569132263
  ,3Daac=60.000001669652114    
  ,3Dlights=CAD
  ]{}{alderson_recenter.u3d}
\caption{\ac{3D} surface scan of the Alderson female phantom performed with the front face camera of a handheld iPhone XS. Note: the figure may appear blank unless opened with Adobe PDF Reader with \url{https://get.adobe.com/flashplayer/} installed.\label{fig:scan3d}}
\end{center}
\end{figure} 
\end{document}